\documentclass[aps,pra,twocolumn,superscriptaddress]{revtex4}
\usepackage{graphicx}
\usepackage{dcolumn}
\usepackage{color}
\usepackage{bm}
\usepackage{amssymb}
\usepackage{amsmath}
\begin{document}
\title{A Floquet formalism for the interaction of magnetically trapped atoms with rf-fields}

\author{A. Chakraborty}
\email[E-mail: ]{carijit@rrcat.gov.in}
\author{S. R. Mishra}
\affiliation{Raja Ramanna Centre for Advanced Technology, Indore-452013, India.}
\affiliation{Homi Bhabha National Institute, Mumbai-400094, India.}

\begin{abstract}
A many mode Floquet theory (MMFT) formalism is applied to study the interaction of a polychromatic rf-field with cold atoms trapped in a quadrupole magnetic trap. In this work, the validity of MMFT approach is first established by comparing its results with those of the previously used formalisms for the cases of single and two frequency rf-fields. We have then used the MMFT formalism to calculate the eigen-energies and transition probabilities for atoms in the quadrupole trap and interacting with a polychromatic rf-field. This composite atom-field system has shown some exquisite features such as lattice like periodic variation in the eigen-energies and large two-photon transition probabilities between the atomic states. This work thus predicts the generation of a lattice type atom trapping potential using polychromatic rf-field, which can be controlled by varying the rf-field parameters.
\end{abstract}
%\pacs{03.75.Be, 37.10.Gh, 05.30.Jp, 67.85.-d, 39.25.+k }
\maketitle

\section{Introduction}
The interaction of an atom with a periodic perturbation has an extensive history \cite{Shirley:1965,Chu:2004}. In the field of ultracold atoms and Bose-Einstein condensation (BEC), oscillating magnetic fields have provided two robust and novel techniques to facilitate the atom trapping. The first one utilizes a low frequency oscillating magnetic field for generating a time orbiting potential (TOP) trap \cite{Minogin:1998}, which provided a simple solution to avoid the Majorana losses at the quadrupole magnetic trap center and enabled the first realization of BEC in a dilute atomic gas \cite{Anderson:1995}. The second technique is based on the proposal by Zobay and Garraway \cite{Zobay:2001,Zobay:2004} which involves the use of a strong high frequency oscillating magnetic field (\textit{i.e.} rf-field) for dressing atomic states in the static magnetic trap. The trapping potential for an atom, generated with the combination of a static magnetic field and a strong dressing rf-field, is known as `rf-dressed potential' \cite{Colombe:2004,White:2006,Lesanovsky:2006:73,Lesanovsky:2006:74,Heathcote:2008,Merloti:R:2013,Morizot:2006,Morizot:2007,Morizot:2008}. These traps, referred as `rf-dressed magnetic traps', are now becoming popular due to their flexibility in tailoring the trapping potential, which is not easily possible with the static magnetic traps alone. The atom trapping potential landscapes and geometries, generated using rf-dressed potentials, can be controlled by varying the rf-field parameters such as amplitude, polarization, frequency and phase \cite{Lesanovsky:2006:73,Lesanovsky:2006:74,Chakraborty:2014}. Using rf-dressed magnetic traps, several atom trapping geometries such as double-well, shell and ring traps have already been demonstrated experimentally \cite{Heathcote:2008,Chakraborty:2016}. This kind of extensive control on trapping and manipulation of cold atoms is useful for various applications including the study of fundamental physics and the matter wave interferometry based precision measurements \cite{Schumm:2005,LeCoq:2006}. 

The progress so far made in this area of research is limited to the use of single \cite{Hofferberth:2006,Morizot:2006,Merloti:R:2013, Chakraborty:2014, Chakraborty:2016} or few frequencies \cite{Alzar:2006,Easwaran:2010,Courteille:2006,Hofferberth:2007,Morgan:e:2010} in the dressing rf-field spectrum. With a polychromatic dressing rf-field and a static magnetic field, the theoretical description of the interaction of an atom with these fields becomes complicated due to the breakdown of the rotating wave approximation (RWA). Therefore accurate prediction of trapping potentials becomes difficult with the existing approaches. Nevertheless, the case of an rf-field having components at two frequencies, one component being stronger than the other, has been treated perturbatively to explain the results of rf-spectroscopy of ultracold atoms trapped in an rf-dressed potential of stronger rf-field \cite{Martin:1988,Easwaran:2010}. For the case of all frequency components in the rf-field being sufficiently strong, a local resonance model was suggested by \citet{Courteille:2006}, which also breaks down for rf-fields with closely spaced frequencies. Hence, there exist a requirement of a more accurate theoretical formalism which can deal with a polychromatic rf-field of arbitrary frequency spectrum and field strength and interacting with atoms trapped in a magnetic trap.

In this article, we have applied a many mode Floquet theory formalism \cite{Ho:1983,Ho:31:1985,Ho:32:1985} to describe the interaction of cold atoms trapped in a magnetic trap with a polychromatic rf-field. This MMFT approach, which is neither perturbative nor based on RWA, seems a prominent formalism to develop an accurate insight into the atom-field interaction problem and calculate the potential landscapes for the trapping of cold atoms. In this approach, the time-dependence in the atom-field interaction Hamiltonian is removed by representing the Hamiltonian in an infinite dimensional Hilbert space composed of the Floquet basis states of the atomic system and the Fourier states of the radiation field. Here we have first compared the MMFT formalism with the earlier established methods for single and two frequency rf-fields. The results have established the superiority of the MMFT approach over the previously used approaches. We have then extended the MMFT formalism to calculate the eigen-energies of the atoms trapped in the quadrupole trap while interacting with the polychromatic rf-field. The results have shown a lattice like periodic structure in the eigen-energy. These lattices are similar to the optical lattices formed by the interfering laser beams, but offer better control over the lattice spacing and depth than the optical lattices. Using the MMFT formalism we have also calculated the time averaged transition probabilities between different atomic states. It is found that there exist a large two-photon transition probability between the atomic states when the rf-field strength is high enough. An appropriate explanation of this high two photon transition probability is presented. 

This article is organized as follows. In section \ref{sec:mmft} the theoretical background of the MMFT has been discussed. In section \ref{subsec:compare}, the results of MMFT approach are compared with those of a dressed state approach under the rotating wave approximation for a single and two mode rf-field. In  section \ref{subsec:poly}, the results of MMFT with a polychromatic rf-field has been discussed in detail. Finally section \ref{sec:concln} presents the conclusion of our work. 

\section{Many mode Floquet theory}\label{sec:mmft}
The many mode Floquet formalism is applied in several areas of science as described in the earlier works \cite{Ho:1983,Ho:1984,Ho:31:1985,Ho:32:1985}. In particular, the use of MMFT is very popular to study interaction of atoms with ultra intense laser fields and to explain several important phenomenon such as multi-photon ionization and generation of higher harmonics \cite{Ruddock:1994,Son:2008}. Here we note that, the interaction of an atom with an rf-field can differ significantly with that of the electric field of a laser radiation due to different nature of interaction. In this article we have applied the MMFT formalism to study the interaction of a cold atom cloud trapped in a quadrupole magnetic trap and interacting with a polychromatic rf-field. 

We consider an arbitrary N-level atomic system interacting with a polychromatic rf-field having an arbitrary polarization. The atomic system is treated quantum mechanically and the polychromatic radiation field ($\textbf{B}(t)$) is treated classically. The evolution of the atomic system can be determined by knowing the interaction Hamiltonian of the time-dependent schr\"{o}dinger equation,
\begin{equation}
i\hbar\frac{\partial\psi}{\partial t}=H\psi,
\end{equation}
where,
\begin{equation}
H=H_0-\boldsymbol{\mu}.\textbf{B}(t),
\end{equation}
with 
\begin{equation}
H_0=\mathcal{H}-\boldsymbol{\mu}.\textbf{B}_s(\textbf{r}).
\end{equation}
Here, $\mathcal{H}$ represents the Hamiltonian without any field, $\boldsymbol{\mu}$ represents the magnetic dipole moment of the atom and $\mathbf{B}_s(\textbf{r})$ represents the static magnetic field of a quadrupole trap. The form of the $\mathbf{B}_s(\textbf{r})$ can be given as,
\begin{equation}\label{eq:Bs}
\textbf{B}_{s}(\textbf{r})=B_q\left(
\begin{matrix}
x\\
y\\
-2z
\end{matrix}
\right),
\end{equation}
where $B_q$ is the radial field gradient of the quadrupole trap. The resulting Larmor frequency $\omega_L$ is then inherently position dependent and can be described as,
\begin{equation}\label{eq:wL}
\omega_L(\textbf{r})=\frac{g_F\mu_B}{\hbar}B_q\sqrt{x^2+y^2+4z^2}.
\end{equation}
The interaction of the polychromatic radiation field $\textbf{B}(t)$ with the above atomic system can be dealt with either by constructing a recursive Floquet Hamiltonian (used in this article) or expanding the complete Hamiltonian as a Magnus series \cite{Klaiber:2009,Butcher:2009}.

The polychromatic rf-field $\textbf{B}(t)$ is composed of several monochromatic field components with their frequencies as $\omega_k$, with $k=0,\pm1,\pm2,...\pm\infty$. Here we consider the frequency spectrum of the polychromatic field to be a comb type in which the width of the individual components $\Delta\omega_k$ is negligible as compared to the separation between the frequency components (\textit{i.e.} $\omega_r$). This field can then be written as,
\begin{equation}\label{eq:brf}
\textbf{B}(t)=\sum_{k} \textbf{B}_{k}\Re\left[e^{i\omega_kt}\right]
\end{equation}
where,
\begin{equation}\label{eq:frequencies}
\omega_k=\omega_0+k\omega_r\ \text{with}\ k=0,\pm 1,\pm 2,...\ ,
\end{equation}
and $\omega_0$ is the frequency of the central component.
Assuming,
\begin{equation}\label{eq:bk}
\textbf{B}_{k}=\textbf{B}_{0}G(\omega_k),
\end{equation}
with $\textbf{B}_{0}$ being the amplitude of the central component and $G(\omega)$ being the envelope function governing the amplitude of the other components relative to the amplitude of the central mode. If we consider only $2N+1$ modes having significant amplitude, then Eq. (\ref{eq:brf}) can be rewritten as, 
\begin{equation}\label{eq:Bosc}
\textbf{B}(t)=\sum_{k=-N}^{k=N} \textbf{B}_{0}G(\omega_k)\Re\left[e^{i\omega_kt}\right].
\end{equation}
Hence the complete system Hamiltonian takes the form,
\begin{equation}\label{eq:finalH}
H=H_0-\sum_{k=-N}^{k=N}\boldsymbol{\mu}.\textbf{B}_{0}G(\omega_k)\Re[e^{i\omega_kt}].
\end{equation}

The Hamiltonian $H_0$ can be represented in the orthonormal atomic basis set $\{|\alpha\rangle\}$ including the Zeemann split levels of the atomic system. The eigen-energy of this Hamiltonian $H_0$ are defined as,
\begin{equation}
\epsilon_\alpha=\langle\alpha|H_0|\alpha\rangle.
\end{equation}
The coupling strength between the different eigen-states due to the time dependent rf-fields is given by,
\begin{equation}
\Omega_{\alpha\beta}^{(k)}=-\frac{1}{2}\langle\alpha|\boldsymbol{\mu}.\textbf{B}_k|\beta\rangle
\end{equation}
where $|\alpha\rangle$ and $|\beta\rangle$ represents different atomic states belonging to the atomic basis set $\{|\alpha\rangle\}$.   

In order to expand the complete Hamiltonian of Eq. (\ref{eq:finalH}), we construct two independent orthonormal fourier basis sets $\{|n\rangle\}$ and $\{|m\rangle\}$ representing the basis frequencies of the polychromatic rf-field, \textit{i.e.} $\omega_0$ and $\omega_r$ respectively. The complete atom-field system now can be described in terms of the composite Floquet basis set,
\begin{equation}\label{eq:fourierbasis}
\{|\alpha nm\rangle\}=\{|\alpha\rangle\}\otimes\{|n\rangle\}\otimes\{|m\rangle\}.
\end{equation}
As the Hilbert space spanned by these Floquet basis sets (Eq. (\ref{eq:fourierbasis})) is infinite dimensional, the Hamiltonian represented in this basis set (called $H^F$) is also an infinite dimensional matrix having the unperturbed Hamiltonian $H_0$ as the central component. We transform the time dependent Hamiltonian of Eq. (\ref{eq:finalH}) to a class of eigen-value problem using the procedure described in an earlier work by \citet{Ho:1983} as,
\begin{equation}\label{eq:HF}
\sum_{\beta}\sum_{n'}\sum_{m'}\langle\alpha nm|H^F|\beta n'm'\rangle\langle\beta n'm'|\epsilon\rangle=\epsilon\langle\alpha nm|\epsilon\rangle
\end{equation}
where $\epsilon$ is the eigen-energy and $|\epsilon\rangle$ is the  eigen-vectors of $H^F$.

The matrix elements of $H^F$ can be determined as,
\begin{multline}
\langle\alpha nm|H^F|\beta n'm'\rangle=H_{\alpha\beta}^{[n-n',m-m']}+\\(n\omega_0+m\omega_r)\delta_{\alpha,\beta}\delta_{n,n'}\delta_{m,m'}
\end{multline}
where $\delta_{ij}$ is the Kronecker delta function. The Hamiltonian matrix elements $H_{\alpha\beta}^{[n-n',m-m']}$ can be written in terms of the unperturbed eigen-energy $\epsilon_\alpha$ and coupling strengths $\Omega_{\alpha\beta}^{(k)}$ as,
\begin{multline}
H_{\alpha\beta}^{[n-n',m-m']}
=\varepsilon_\alpha\delta_{\alpha,\beta}\delta_{n,n'}\delta_{m,m'}+\\ \sum_{k=-N}^{N}\Omega_{\alpha\beta}^{(k)}\left(\delta_{n+1,n'}\delta_{m+k,m'}+\delta_{n-1,n'}\delta_{m-k,m'}\right).
\end{multline}
Hence, by determining the matrix element $H_{\alpha\beta}^{[n-n',m-m']}$, we can determine the complete infinite dimensional matrix $\mathbb{H}^F_{\omega_k}$. The suffix $\omega_k$ in the Hamiltonian denotes the inclusion of all the components of the polychromatic rf-field with frequencies $\omega_k$. Matrix containing only the contribution of $\omega_0$, denoted as $\mathbb{H}^F_{\omega_0}$, is also an infinite dimensional matrix, and it is a sub-matrix of the composite Floquet matrix $\mathbb{H}^F_{\omega_k}$. An examination of the expanded form of these Floquet matrices can provide enough insight about their recursive and composite structure. Central region of these infinite dimensional matrices are shown below,
\begin{widetext}
\begin{equation}
\mathbb{H}^F_{\omega_k}=\left[
\begin{matrix}
\ddots                    & \vdots                     & \                & \      & \              & \vdots & \ \\ 
\cdots                    & \mathbb{H}^F_{\omega_0}+2\hbar\omega_r\mathbb{I}  & \mathbb{Q}_1              & \mathbb{Q}_2    & \mathbb{Q}_3              & \mathbb{Q}_4               & \cdots \\
\                         & \mathbb{Q}_1^T             & \mathbb{H}^F_{\omega_0}+\hbar\omega_r \mathbb{I} & \mathbb{Q}_1    & \mathbb{Q}_2              & \mathbb{Q}_3               & \ \\
\                         & \mathbb{Q}_2^T             & \mathbb{Q}_1^T            & \mathbb{H}^F_{\omega_0}      & \mathbb{Q}_1              & \mathbb{Q}_2               & \ \\
\                         & \mathbb{Q}_3^T             & \mathbb{Q}_2^T            & \mathbb{Q}_1^T  &             \mathbb{H}^F_{\omega_0}-\hbar\omega_r\mathbb{I}  & \mathbb{Q}_1               & \ \\
\cdots                    & \mathbb{Q}_4^T             & \mathbb{Q}_3^T            & \mathbb{Q}_2^T  &          \mathbb{Q}_1^T            & \mathbb{H}^F_{\omega_0}-2\hbar\omega_r\mathbb{I}  & \cdots \\ 
\                         & \vdots                     & \                & \      & \              & \vdots & \ddots \\ 
\end{matrix}\right];
\end{equation}
\begin{tabular}{cc}
\begin{minipage}{3cm}
\begin{equation*}
\mathbb{H}^F_{\omega_0}=\left[
\begin{matrix}
\ddots           & \vdots                        & \                & \       & \            & \vdots       & \ \\ 
\cdots           & \mathbb{H}_0+2\hbar\omega_0\mathbb{I}   & \mathbb{Y}_0     & \mathbb{O}       & \mathbb{O}            & \mathbb{O}            & \cdots \\
\                & \mathbb{Y}_0                  & \mathbb{H}_0+\hbar\omega_0\mathbb{I} & \mathbb{Y}_0 & \mathbb{O}      & \mathbb{O}   & \ \\
\       & \mathbb{O}      & \mathbb{Y}_0                  & \mathbb{H}_0               & \mathbb{Y}_0          & \mathbb{O}   & \ \\
\       & \mathbb{O}      & \mathbb{O}            & \mathbb{Y}_0   & \mathbb{H}_0-\hbar\omega_0\mathbb{I} & \mathbb{Y}_0                & \ \\
\cdots  & \mathbb{O}      & \mathbb{O}            & \mathbb{O}              & \mathbb{Y}_0               & \mathbb{H}_0-2\hbar\omega_0\mathbb{I} & \cdots \\ 
\                & \vdots                        & \                & \       & \            & \vdots       & \ddots \\ 
\end{matrix}\right];
\end{equation*}
\end{minipage}
\hspace{7.8cm}
\begin{minipage}{3cm}
\begin{equation*}
\mathbb{Q}_k=\left[
\begin{matrix}
\ddots           & \vdots           & \                & \            & \               & \vdots          & \ \\ 
\cdots           & \mathbb{O}       & \mathbb{Y}_k     & \mathbb{O}   & \mathbb{O}      & \mathbb{O}      & \cdots \\
\                & \mathbb{Y}_{-k}  & \mathbb{O}       & \mathbb{Y}_k & \mathbb{O}      & \mathbb{O}      & \ \\
\                & \mathbb{O}       & \mathbb{Y}_{-k}  & \mathbb{O}   & \mathbb{Y}_k    & \mathbb{O}      & \ \\
\                & \mathbb{O}       & \mathbb{O}       & \mathbb{Y}_{-k} & \mathbb{O}   & \mathbb{Y}_k    & \ \\
\cdots           & \mathbb{O}       & \mathbb{O}       & \mathbb{O}      & \mathbb{Y}_{-k} & \mathbb{O}   & \cdots \\ 
\                & \vdots           & \                & \            & \               & \vdots          & \ddots \\ 
\end{matrix}\right]
\end{equation*} 
\end{minipage}
\end{tabular}
\begin{tabular}{cccc}
\begin{minipage}{3cm}
\begin{equation*}
\mathbb{Y}_k=\left(
\begin{matrix}
0                      & \Omega_{\alpha\beta}^{(k)}   & \Omega_{\alpha\gamma}^{(k)}   & \cdots   \\
\Omega_{\beta\alpha}^{(k)}  & 0                       & \Omega_{\beta\gamma}^{(k)}    & \        \\
\Omega_{\gamma\alpha}^{(k)} & \Omega_{\beta\gamma}^{(k)}   & 0                        & \        \\
\vdots                 & \                       & \                        & \ddots   \\ 
\end{matrix}\right);
\end{equation*}
\end{minipage}
\hspace{2.5cm}
\begin{minipage}{3cm}
\begin{equation*}
\mathbb{O}=\left(
\begin{matrix}
0         & 0  & 0     & \cdots   \\
0         & 0  & 0     & \        \\
0         & 0  & 0     & \        \\
\vdots    & \  & \     & \ddots   \\ 
\end{matrix}\right);
\end{equation*}
\end{minipage}
\hspace{0.8cm}
\begin{minipage}{3cm}
\begin{equation*}
\mathbb{H}_0=\left(
\begin{matrix}
\varepsilon_{\alpha}   & 0                    & 0                        & \cdots   \\
0                      & \varepsilon_{\beta}  & 0                        & \        \\
0                      & 0                    & \varepsilon_{\gamma}     & \        \\
\vdots                 & \                    & \                        & \ddots   \\ 
\end{matrix}\right);
\end{equation*}
\end{minipage}
\hspace{0.9cm}
\begin{minipage}{4cm}
\begin{equation*}
\mathbb{I}=\left(
\begin{matrix}
1         & 0  & 0     & \cdots   \\
0         & 1  & 0     & \        \\
0         & 0  & 1     & \        \\
\vdots    & \  & \     & \ddots   \\ 
\end{matrix}\right).
\end{equation*}
\end{minipage}
\end{tabular}
\end{widetext}

As the eigen-value problem of Eq. (\ref{eq:HF}) cannot be solved numerically with the infinite dimensional Floquet matrices $\mathbb{H}_{\omega_k}^{F}$ and $\mathbb{H}_{\omega_0}^{F}$, due to practical limitations, these matrices should be truncated after a particular order $N$ and $N_o$ respectively. The eigen-energies of the complete matrix $\mathbb{H}_{\omega_k}^{F}$ are also called `quasi-energies' as they contain contributions from all the Floquet modes $n$ and $m$.

In terms of the eigen-vectors of the Hamiltonian matrix $\mathbb{H}_{\omega_k}^F$, denoted as $|\epsilon nm\rangle$, we can obtain the time averaged transition probabilities between any two atomic states $|\alpha\rangle$ and $|\beta\rangle$ as,
\begin{equation}
\bar{P}_{\alpha\rightarrow\beta}=\sum_{n,m}\sum_{\epsilon n'm'}|\langle\beta nm|\epsilon n'm'\rangle\langle\epsilon n'm'|\alpha 00\rangle|^2.
\end{equation}
These time-averaged transition probabilities are useful in determination of the `avoided crossings' and probability of the trapping of the atom. Here, we note that any avoided crossing is usually characterized by a different non-adiabatic transition probability, called Landau-Zener transition probability $P_{LZ}$ \cite{Breuer:11:1989,Breuer:140:1989}, originally proposed by Landau and Zener for two level systems and later modified by \citet{Vitanov:1997} for a multi-level system. At any avoided crossing, $P_{LZ}$ can be determined in terms of the energy gap at the crossing $\delta\epsilon$, spread of the avoided crossing in frequency space $\delta\omega$, and rate of change of the frequency with time $\dot{\omega}$ which is dependent on velocity of the atom with which it traverses the avoided crossing. This is given as,
\begin{equation}
P_{LZ}=\exp{\left(-\frac{\pi}{2}\frac{\delta\epsilon\delta\omega}{\dot{\omega}}\right)}.
\end{equation}
Due to the induction of avoided crossing in all the resonant interactions between the atom and radiation field, transition probability $P_{LZ}$ determines the validity of the adiabaticity condition in the system.

\section{Results and discussion}\label{sec:results}
In order to conduct a thorough numerical study, objective of which is to provide physical parameters of experimental interest, we consider the case of $^{87}Rb$ atom in the hyperfine state $|F=2\rangle$, which is interacting with an rf-field in presence of a static magnetic field. In the following subsection, first we compare the results of the MMFT approach with those of the earlier approaches \cite{Chakraborty:2014,Courteille:2006} for single and two frequency rf-fields. Further, we have also scrutinized the effect of the truncation of the infinite Floquet matrices $\mathbb{H}_{\omega_k}^{F}$, $\mathbb{H}_{\omega_0}^{F}$ and $\mathbb{Q}_k$. 

\subsection{Comparison of MMFT with other approaches}\label{subsec:compare}
To begin with, we describe the interaction of the atoms with single frequency rf-field in presence of static magnetic field using an approach based on the rotating wave approximation in dressed state picture (RWDS) and compare the results with MMFT formalism. In RWDS approach, the counter rotating terms are removed from the Hamiltonian for analytical simplicity. 

We consider an atom trapped in a static quadrupole field $\textbf{B}_s(\textbf{r})$ given by the Eq. (\ref{eq:Bs}) and exposed to a monochromatic rf-field with frequency $\omega_0$ and of form 
\begin{equation}\label{eq:Brf}
\textbf{B}_{0}(t)=\left(
\begin{matrix}
B^{x}_{0} cos(\omega_0 t)\\
B^{y}_{0} cos(\omega_0 t-\phi_y)\\
B^{z}_{0} cos(\omega_0 t-\phi_z)
\end{matrix}
\right),
\end{equation}
where $B^{x}_0$, $B^{y}_0$ and $B^{z}_0$ are the rf-field strengths along the Cartesian axes and $\phi_y$, $\phi_z$ are the phase differences between the x-y and x-z component respectively. To remove the time dependence, we transform the composite atom-field system Hamiltonian by applying an unitary transformation and under the rotating wave approximation the transformed Hamiltonian becomes,
\begin{equation}\label{eq:H}
H=\frac{g_F\mu_B}{\hbar}\left[F_3\left(|\textbf{B}_{s}\textbf{(r)}|+B_{0}^{||}\right)+\sum_{i=1,2}F_iB_{0}^{\bot i}\right].
\end{equation}
Here $B_{0}^{\perp 1}$, $B_{0}^{\perp 2}$ and $B^{||}_0$ are amplitudes of the perpendicular and parallel components of the applied rf-field $\textbf{B}_0(t)$ with respect to the local static field vector $\textbf{B}_s(\textbf{r})$. Due to the vectorial nature of coupling between the atomic hyperfine sub-levels and the applied rf-field, only the perpendicular components contribute to the final coupling strength $\Omega_0$, and the contribution from the parallel component can be neglected using the adiabatic approximation. Using this Hamiltonian in Eq. (\ref{eq:H}), the eigen-energies for different Zeeman hyperfine sub-levels ($m_F=-F$ to $F$) can be obtained in RWDS approach in terms of the detuning $\delta_0$ and the coupling strength or Rabi frequency $\Omega_0$ as,
\begin{equation}\label{eq:pot_compact}
\epsilon_{0}(\textbf{r})=m_F\hbar\sqrt{\delta_0^2+|\Omega_0|^2},
\end{equation}
with
\begin{equation}\label{eq:delta}
\delta_0(\textbf{r})=\omega_L(\textbf{r})-\omega_0,
\end{equation}
\footnotesize\begin{equation}\label{eq:omega}
|\Omega_0|^2=\left(\frac{g_F\mu_B}{2\hbar}\right)^2{\left[2B_{0}^{\bot 1} B_{0}^{\bot 2}\sin\gamma+\sum_{i=1,2}(B_{0}^{\bot i})^2\right]},
\end{equation}\normalsize
where $\gamma$ is the phase separation between the perpendicular components $B_{0}^{\perp 1}$ and $B_{0}^{\perp 2}$ \cite{Chakraborty:2014}.

\begin{figure}[t]
\includegraphics[width=8.5 cm]{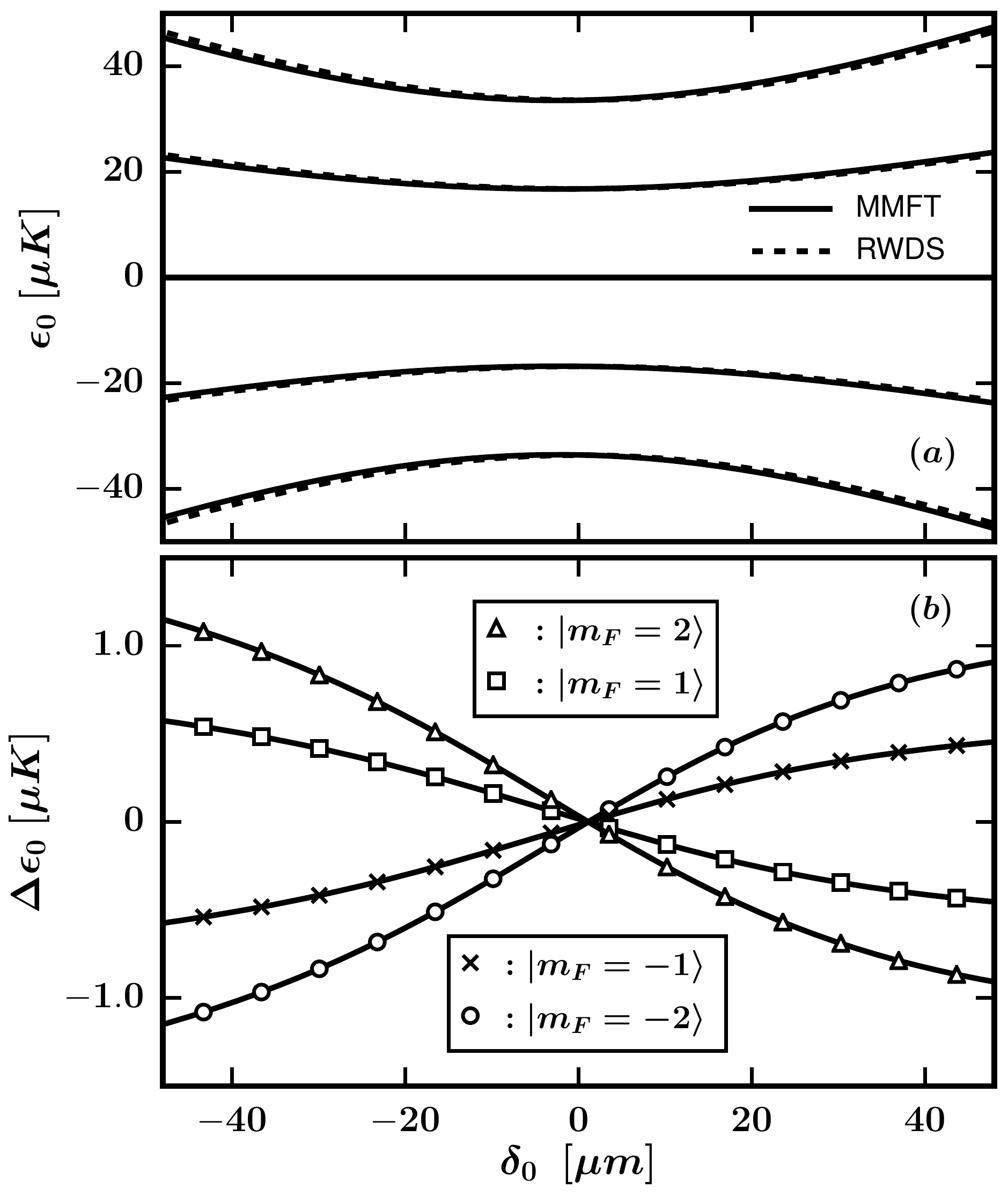}
\caption{\label{fig:comparison} Comparison of the MMFT and RWDS method for atoms interacting with a single frequency rf-field. (a) Variation of eigen-energies ($\epsilon_0$) for different Zeeman hyperfine sub-levels as a function of $\delta_0$, using the RWDS and MMFT (for truncation order  $N_o=5$). (b) Difference between the predicted energy values $(\Delta\epsilon_0)$ as a function of $\delta_0$ for above two methods. The other parameters are $\omega_0=2\pi\times2$ MHz, $B_0^x=1$ G, $B_0^y=B_0^z=0$, $B_q=100\ G\ cm^{-1}$.}
\end{figure}

Figure \ref{fig:comparison} (a) shows the eigen-energies $\epsilon_0(\textbf{r})$ seen by the atoms of $^{87}Rb$ for different Zeeman hyperfine sub-levels of the $|F=2\rangle$ state under RWDS and MMFT approaches as a function of detuning $\delta_0$. Close agreement (qualitative and quantitative) between the results from both the approaches ensures the validity of the many mode Floquet theory to predict the eigen-energies of an atomic system interacting with a single frequency rf-field. To correlate the frequency detuning with the spatial scale Eq. (\ref{eq:delta}) has been used along with Eq. (\ref{eq:wL}) with the radial field gradient $B_q$ = 100 $G \ cm^{-1}$.

\begin{figure}[t]
\includegraphics[width=8.5 cm]{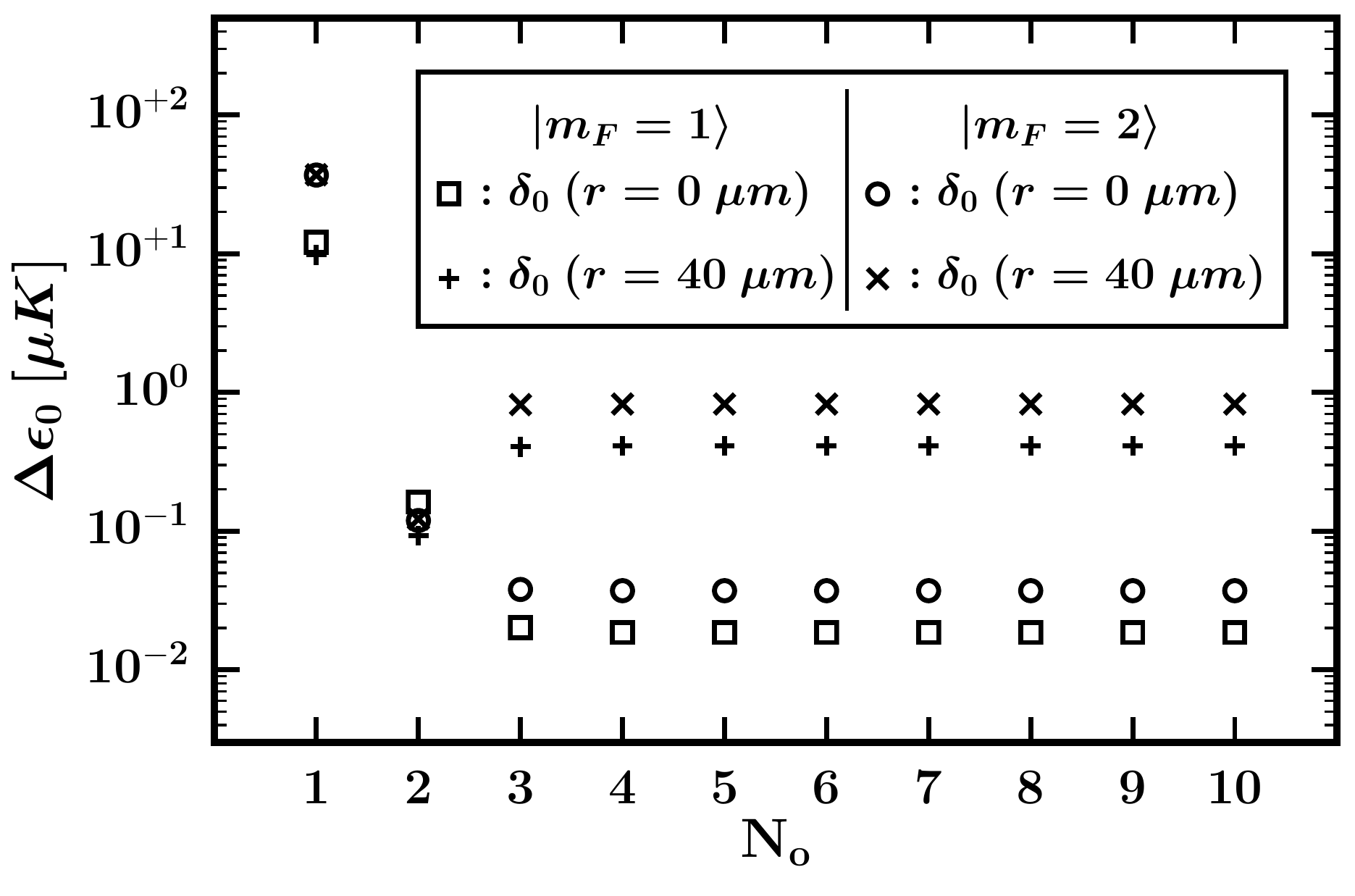}
\caption{\label{fig:Ni} Calculated energy difference $(\Delta\epsilon_0)$ between the RWDS and MMFT methods as a function of truncation order $N_o$ in the infinite matrix $\mathbb{H}^F_{\omega_0}$ for two different values of detuning $\delta_0\left({r=0\ \mu m}\right)$ and $\delta_0\left({r=40\ \mu m}\right)$. The saturation of energy difference above $N_{o}$ = 3 provides the minimum order of truncation to be used. The other parameters are $\omega_0=2\pi\times2$ MHz, $B_0^x=1$ G, $B_0^y=B_0^z=0$, $B_q=100\ G\ cm^{-1}$.}
\end{figure}

The energy difference $\Delta\epsilon_0=\epsilon_0^{RWDS}-\epsilon_0^{MMFT}$ is shown in Fig. \ref{fig:comparison} (b) as a function of detuning. Perfect agreement between energy values from two approaches at zero detuning proves the excellent validity of both the approaches at the resonance point. The minuscule difference between the two approaches around the resonance may be due to the removal of the counter rotating term from the RWDS Hamiltonian under the rotating wave approximation. The effect of truncation order $N_o$ is studied by comparing the difference between the eigen-energies of RWDS and MMFT approaches \textit{i.e.} $\Delta\epsilon_0$ as a function of $N_o$. The results are plotted in Fig. \ref{fig:Ni} for two different detuning values $\delta_0\left({r=0\ \mu m}\right)$ and $\delta_0\left({r=40\ \mu m}\right)$. This suggests a minimum value of $N_o$, \textit{i.e.} $N_o>3$, is sufficient to minimize the truncation error in the MMFT approach. 

To make a further comparison of the MMFT and RWDS approaches, we consider the case of an rf-field having two frequency components. The second component of the rf-field, \textit{i.e.} $\textbf{B}_1(t)$, has frequency $\omega_1$ and coupling strength $|\Omega_1|$, with other parameters like polarization considered same as $\textbf{B}_0(t)$. In RWDS approach, as used earlier \cite{Easwaran:2010}, the ratio of the coupling strengths is assumed to be `relatively small' $(\Omega_1/\Omega_0\rightarrow 0)$, and hence the interaction of the second component of the rf-field can be treated perturbatively where the eigen-energy values are considered same as $\epsilon_0$ . To compare the results of this approach with the results of MMFT, we have compared the eigen-energies obtained by MMFT using $\mathbb{H}_{\omega_k}^{F}$ for $k=1$. An appropriate amplitude envelope $G(\omega)$ was used, which has a variable strength of the positive component $\omega_{k=+1}$ while having the strength of the negative component $\omega_{k=-1}$ zero.

\begin{figure}[t]
\includegraphics[width=8.5 cm]{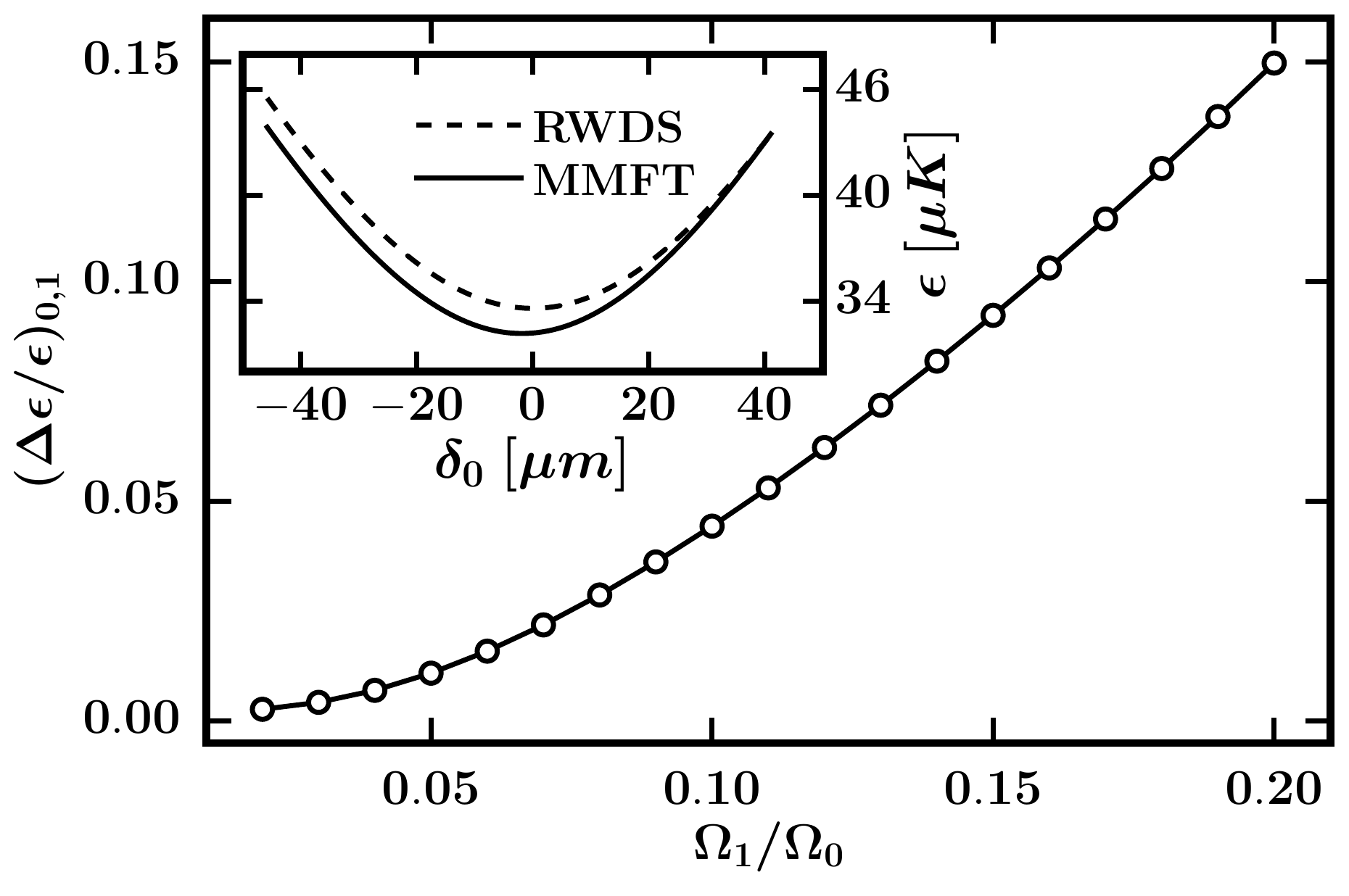}
\caption{\label{fig:strlimit} Relative shift in energy $(\Delta\epsilon/\epsilon)_{0,1}$ as a function of relative strength $(\Omega_1/\Omega_0)$. In the inset, eigen-energies ($\epsilon_{0,1}$) obtained from the RWDS (dashed curve) and the MMFT (solid curve) for relative strength $\Omega_1/\Omega_0\ =\ 0.1$ is shown as a function of detuning $\delta_0$. The other parameters are $\omega_0=2\pi\times2$ MHz, $\omega_1=2\pi\times1.5$ MHz, $B_0^x=1$ G, $B_0^y=B_0^z=0$, $B_q=100\ G\ cm^{-1}$, $N_o=5$.}
\end{figure}

Figure \ref{fig:strlimit} shows the dependence of the relative energy difference between the eigen-energies obtained from both the RWDS and MMFT method as a function of the relative coupling strength $\Omega_1/\Omega_0$, where the relative energy difference has been described as,
\begin{equation}
\left(\frac{\Delta\epsilon}{\epsilon}\right)_{0,1}=\left(\epsilon_0^{RWDS}-\epsilon_{0,1}^{MMFT}\right)/\epsilon_{0,1}^{MMFT}.
\end{equation}
The monotonic increase in the energy shift shows the limitation of the perturbative approach. For example, with $\Omega_1$ $\simeq 20\%$ of $\Omega_0$, the difference in the eigen-energy is considerably large ($15\%$). The resulting eigen-energies for the two methods have also been plotted as a function of detuning in the inset of Fig. \ref{fig:strlimit} to bring out the difference clearly.

Since the perturbative approach shows very large shifts in the eigen-energies even for small relative coupling strengths, interaction of an atom with two rf-fields with comparable strength ($\Omega_1\simeq\Omega_0$) can not be treated with this approach. In a work by \citet{Courteille:2006} this atom-field composite system has been treated with a local resonance method (LRM) in which the atom is considered interacting with only a single frequency component of the rf-field (having lowest detuning) at any particular trap position (\textbf{r}). The effect of the off-resonant components of the rf-field are incorporated by evaluating the ac stark-shifts $\Gamma(\textbf{r})$ and modifying the dressed state eigen-energies as discussed below. To obtain the eigen-energies of this system, we determine the stark shifts for the off resonant rf-fields as a function of detuning (\textit{i.e.} position) by using the relation,
\begin{subequations}
\begin{eqnarray}
\Gamma_0(\textbf{r})=\frac{\Omega_0^2}{\delta_1(\textbf{r})},\\ 
\Gamma_1(\textbf{r})=\frac{\Omega_1^2}{\delta_0(\textbf{r})},
\end{eqnarray}
\end{subequations}
where $\delta_0$ and $\delta_1$ are the detunings respective to the frequency components $\omega_0$ and $\omega_1$. After considering these stark shifts, the eigen-energies corresponding to the two frequency components becomes,
\begin{subequations}
\begin{eqnarray}
\epsilon_0(\textbf{r})=\sqrt{\left(\delta_0(\textbf{r})+2\Gamma_0(\textbf{r})\right)^2+|\Omega_0|^2},\\ 
\epsilon_1(\textbf{r})=\sqrt{\left(\delta_1(\textbf{r})+2\Gamma_1(\textbf{r})\right)^2+|\Omega_1|^2}.
\end{eqnarray}
\end{subequations}
The final eigen-energy of the two component atom-field composite system can be obtained using the LRM approach as,
\footnotesize \begin{equation}
\epsilon_{0,1}(\textbf{r})=m_F\hbar\Bigg[\Theta\Big[\Delta\delta(\textbf{r})\Big]\epsilon_0(\textbf{r})+\Theta\Big[-\Delta\delta(\textbf{r})\Big]\epsilon_1(\textbf{r})\Bigg]
\end{equation}\normalsize
where $\Delta\delta(\textbf{r})=|\delta_1(\textbf{r})|-|\delta_0(\textbf{r})|$, determines the proximity of resonance due to one component over the other component. The Heaviside step function $\Theta[x]$ facilitates the transition of the final eigen-energy $\epsilon_{0,1}(\textbf{r})$ from one eigen-energy to another at the resonance crossing points $r_c$, defined by $\Delta\delta(r_c)=0$ (\textit{i.e.} $\delta_0(r_c)=\delta_1(r_c)$). The key approximation in this method is the separation between the two frequencies (\textit{i.e.} $\Delta\omega=\omega_1-\omega_0$) being sufficient so that the jump in energy at $r_c$ is minimum. Hence, to obtain the validity regime of the LRM approach, we define the dimensionless parameter $\zeta$ as,
\begin{equation}
\zeta=\displaystyle{\lim_{r\to r_c}}\frac{\epsilon_0(\delta_0=0)-\epsilon_1(\delta_1=0)}{\epsilon_0(r)-\epsilon_1(r)}.
\end{equation}

\begin{figure}[t]
\includegraphics[width=8.5 cm]{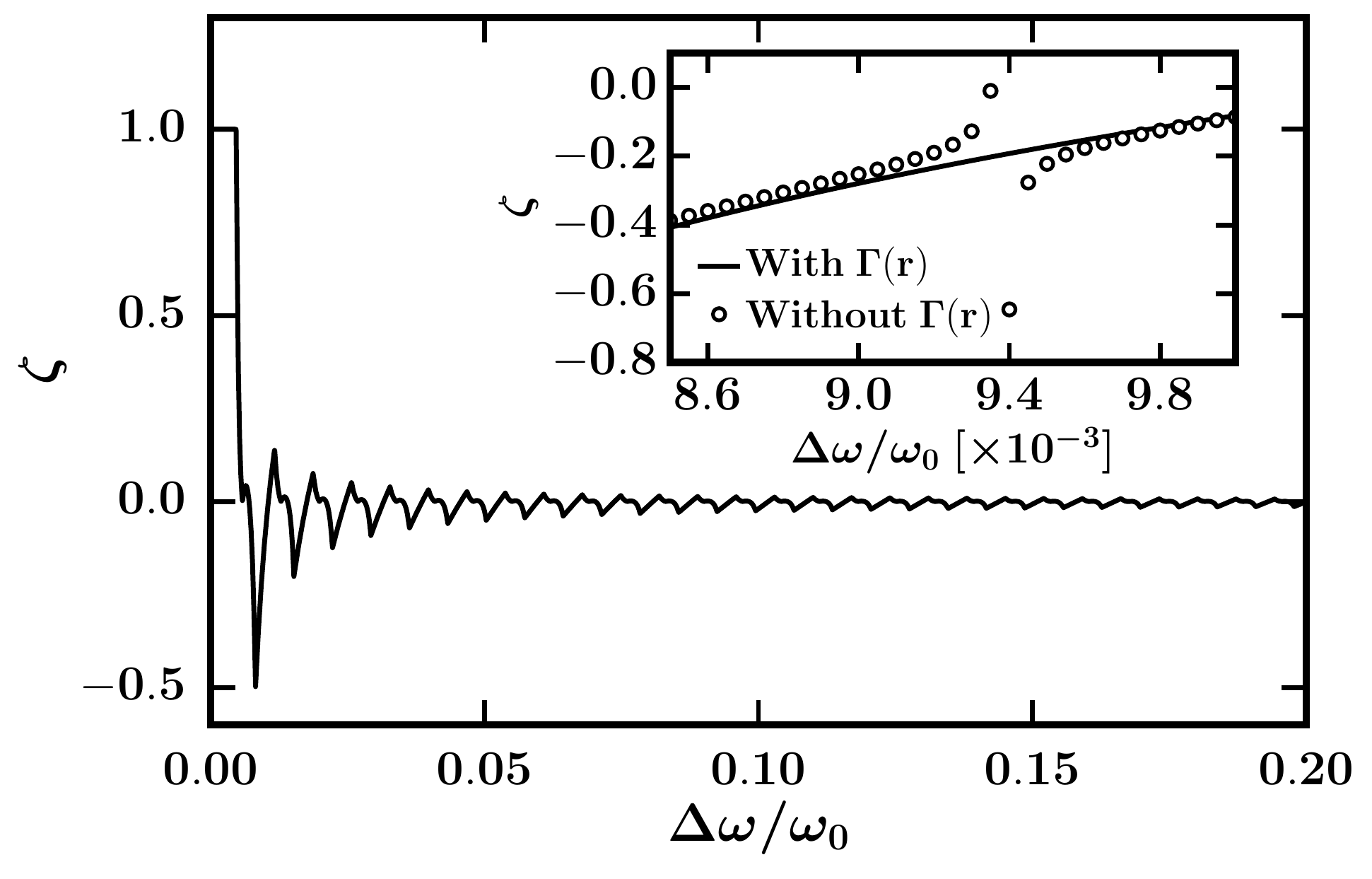}
\caption{\label{fig:dwlimit} Dimensionless quality parameter $\zeta$ for equally strong rf-fields (\textit{i.e.} $|\Omega_0|^2\ =\ |\Omega_1|^2$) as a function of relative frequency separation $\Delta\omega/\omega_0$ in the LRM approach. The inset shows the magnified plot of a particular region with (continuous curve) and without (hollow circles) considering the stark shifts $\Gamma(\textbf{r})$ due to the off-resonant rf-fields. Here $\omega_0=2\pi\times2$ MHz, $B_0^x=1$ G, $B_0^y=B_0^z=0$ and $B_1^x=1$ G, $B_1^y=B_1^z=0$, $B_q=100\ Gcm^{-1}$.}
\end{figure}

Figure \ref{fig:dwlimit} shows the plot of the dimensionless parameter $\zeta$ as a function of the relative frequency separation $\Delta\omega/\omega_0$. At large frequency separations, the transition from one eigen-energy to another one at the resonance crossing points ($r_c$) becomes smooth while the depth of the potential minimum at the resonance points ($\delta_0=\delta_1=0$) becomes equal \cite{Courteille:2006}. This leads $\zeta$ to be nearly zero at the large $\Delta\omega/\omega_0$. On the contrary at the smaller frequency separations, both the shift in the eigen-energy minima and the energy-shift at the resonance crossing points become finite. This leads to a large increase in the $\zeta$ value which questions the validity of the LRM approach at the limit $\Delta\omega/\omega_0\rightarrow 0$. 

The effect of inclusion of the stark-shifts due to the non-resonant rf-fields in the LRM approach is shown in the inset of Fig. \ref{fig:dwlimit}. The smooth solid line shows the value of $\zeta$ while keeping the contribution of the stark-shifts, whereas the hollow circles shows the values without considering the stark-shifts. It is clear from this plot that the inclusion of stark-shift avoided the sudden divergences and provided a smooth variation of $\zeta$.   

The outcome of this subsection is to establish that even for only two rf-fields with relative strength more than 0.1 and relative frequency separation less than 0.05, both the earlier approaches RWDS and LRM break down. Therefore, it is natural to look for a different formalism, such as MMFT, to cover these regimes of the parameters where the earlier approaches are inadequate. The MMFT approach, being neither perturbative nor based on RWA, provides nearly accurate eigen-energies using appropriate truncation order. Therefore, in the next subsection we will employ the MMFT method to solve the interaction of an atomic system with a polychromatic rf-field having more than two frequency components.

\subsection{MMFT with polychromatic rf-fields}\label{subsec:poly}

In this subsection, we discuss the results of MMFT formalism for the interaction of magnetically trapped atom cloud with a polychromatic rf-field as described by Eq. (\ref{eq:Bosc}). We consider the interaction of an atomic system of $^{87}Rb$ in $|F=2\rangle$ hyperfine state with an rf-field having the comb frequency spectrum given by Eq. (\ref{eq:Bosc}) with $N=4$. The eigen-energies ($\epsilon$) obtained from the sparse Floquet Hamiltonian, $\mathbb{H}_{\omega_k}^{F}$, are periodic in energy because of the contribution of the Fourier components in the basis set $\{|\alpha nm\rangle\}$. Hence, the eigen-energy spectrum can be bundled into Floquet manifolds by the slow running basis index $n$. The zeroth manifold (\textit{i.e.} with $\epsilon_{\alpha,n=0,m}$) lies closest to the adiabatic iso-energy surfaces and has been studied extensively in this subsection. The most important feature of the eigen-energies $\epsilon_{\alpha,0,m}$ is their spatial periodicity. This spatial periodicity is dependent on the separation between the field modes and can be well described in terms of a dimensionless parameter $\xi=(\omega_L-\omega_0)/\omega_r$. In this new scale, we can expect the periodic eigen-energy minimum to be present at integer values of $\xi$, for an arbitrary separation frequency $\omega_r$.

\begin{figure}[t]
\includegraphics[width=8.5 cm]{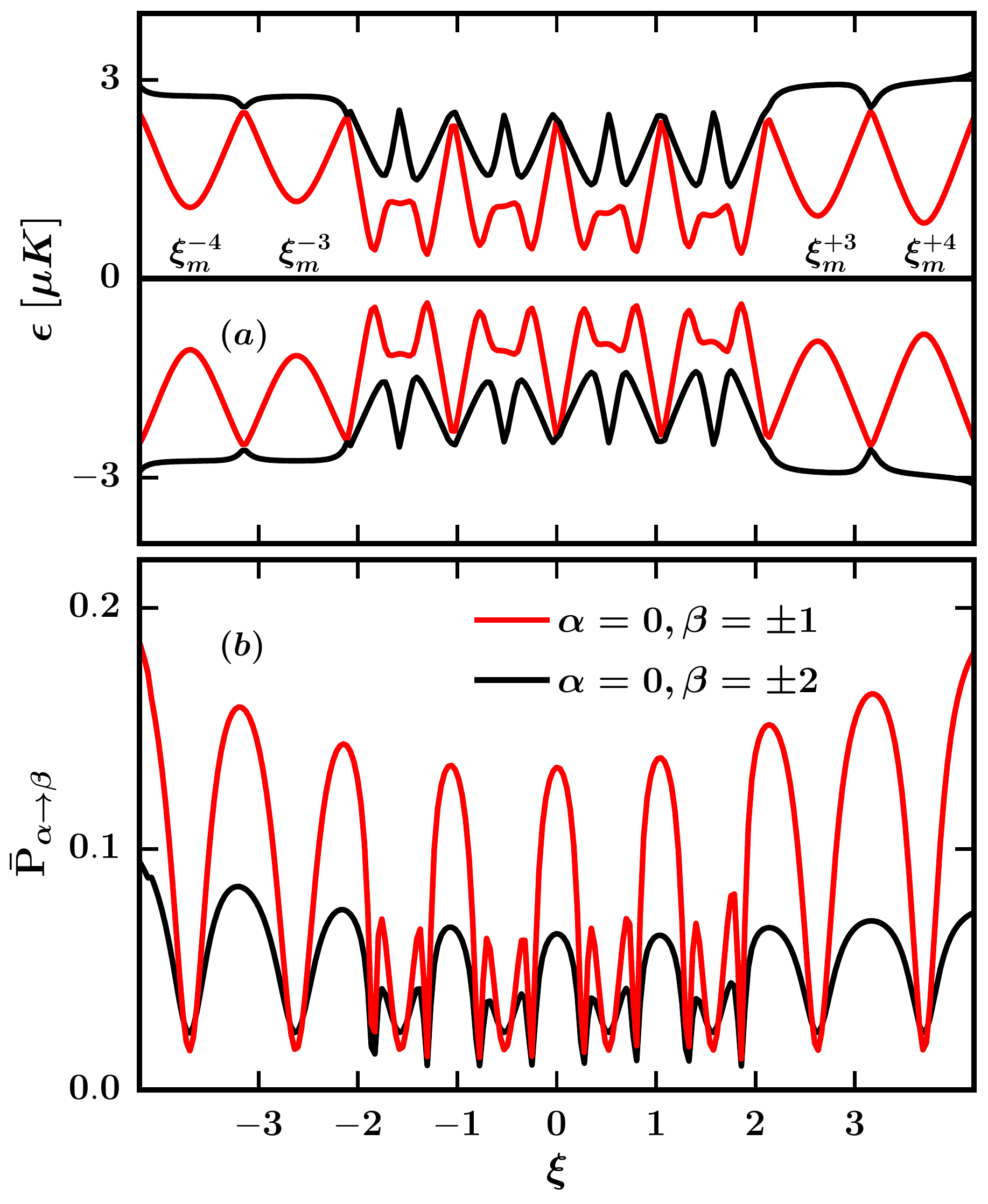}
\caption{\label{fig:ep}(a) Variation in eigen-energies $(\epsilon)$ as a function of dimensionless parameter $\xi=(\omega_L-\omega_0)/\omega_r$ for $N=4$. Here, $\xi_m^{k}$ correspond the eigen-energy minimum for the $k^{th}$ component of the rf-field. (b) Time averaged transition probabilities $\left(\bar{P}_{\alpha\rightarrow\beta}\right)$ between the $|0\rangle$ and $|\pm 1\rangle$, $|\pm 2\rangle$ states as a function of $\xi$. The other parameters are $\omega_0=2\pi\times\ 2$ MHz, $\omega_r=2\pi\times\ 100$ kHz and $B_0^x=1$ G, $B_0^y=B_0^z=0$, $B_q=100\ G\ cm^{-1}$, $N_o=5$.}
\end{figure}

The central region of the zeroth manifold of the Floquet spectrum has been plotted in Fig. \ref{fig:ep} (a) as a function of $\xi$. This plot shows a plethora of avoided crossings in the eigen-energy structure. As expected, the number of avoided crossings across the zero energy is proportional to the number of side frequency modes in the rf-field, which is 8 (for $k=\pm1\to\pm4$). Therefore, the field mode index $k$ can be used to mark the position of the local eigen-energy minimum as $\xi_m^k$. In the plot depicted, there are four single minima at the edge of the spectrum (marked as $\xi_m^{\pm3}$ and $\xi_m^{\pm4}$) and four minima which are split into eight structures at the central region. This additional splitting of the eigen-energy minima in the central region is due to the additional avoided crossings which are generated across the lower (solid red) and upper (solid black) eigen-energy curves. Due to the comparatively large separations between the avoided crossings, the LZ-transition probability $P_{LZ}$ for these wells is $\sim 1.0$ for an atom cloud of temperature $\sim\ 10 \mu K$. This high value of $P_{LZ}$ is one of the key requirements for  atom trapping applications. Hence, this multi-well energy structure shown in Fig. \ref{fig:ep} can be utilized as a lattice potential for atom trapping, similar to the one obtained using interfering counter propagating laser beams (\textit{i.e.} optical lattices). Due to the dependence of the distance between the wells on the separation frequency $\omega_r$, this trap offers more control and can be manoeuvred more easily than the optical lattices \cite{Wright:2000}.

The other quantities of interest are the time averaged transition probabilities $(\bar{P}_{\alpha\rightarrow\beta})$ for single $(|0\rangle\rightarrow|\pm 1\rangle)$ and two $(|0\rangle\rightarrow|\pm 2\rangle)$ photon transitions, which are plotted as a function of $\xi$ in Fig. \ref{fig:ep} (b). The occurrence of the minimum in the transition probability in Fig. \ref{fig:ep}(b) at specific $\xi$ values which correspond to the weakly split minimum in the eigen-energy value (Fig. \ref{fig:ep}(a)), establishes the existence of the avoided crossings at those values of $\xi$. One of the prime feature of this plot (Fig. \ref{fig:ep}(b)) is the comparable strength of both (single and two photon) transition probabilities. Such large two photon transition probabilities are rare in the optical domain because of the weaker atom-field interactions as compared to the atom-field interaction in radio frequency regime. Due to the presence of the multiple frequencies of rf-field, there can be several combinations of frequencies, such that summation of frequencies in a given combination equals to the two photon transition frequency. 

\begin{figure}[t]
\includegraphics[width=8.5 cm]{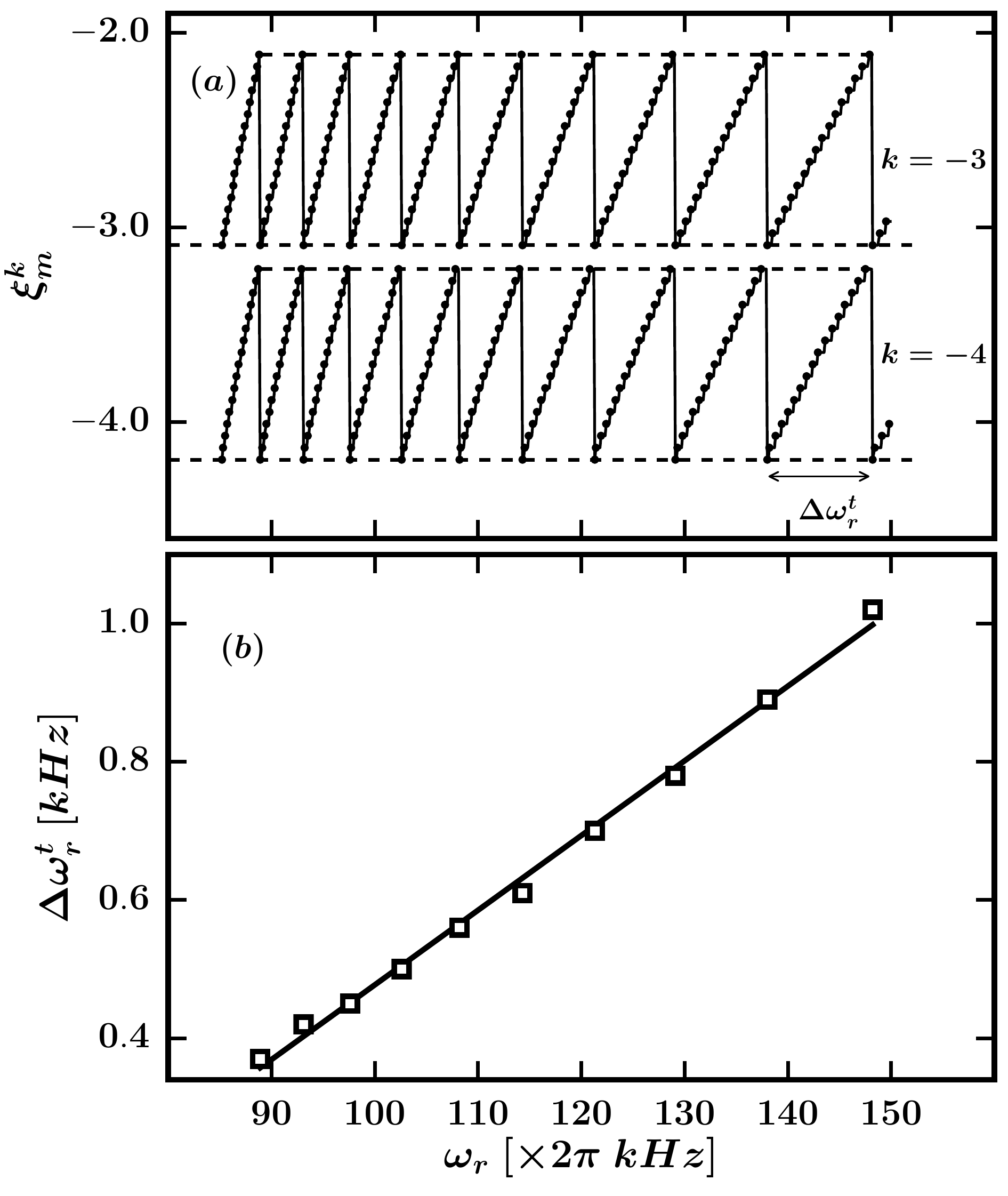}
\caption{\label{fig:wr} (a) Variation in position of local eigen-energy minimum $\xi_m^k$ ($\xi_m^{-3}$ and $\xi_m^{-4}$) as a function of $\omega_r$. The dashed lines show the lower and upper ends of $\xi_m^k$ values. (b) Variation in the separation between the terminal frequencies ($\omega_r^t$), \textit{i.e.} $\Delta\omega_r^t$, as a function of the separation frequency $\omega_r$ (open square). The continuous line is a linear fit. The other parameters are $\omega_0=2\pi\times\ 2$ MHz, $B_0^x=1$ G, $B_0^y=B_0^z=0$ and $B_q=100\ G\ cm^{-1}$, $N_o=5$, $N=4$.}
\end{figure}

\begin{figure*}[t]
\includegraphics[width=17 cm]{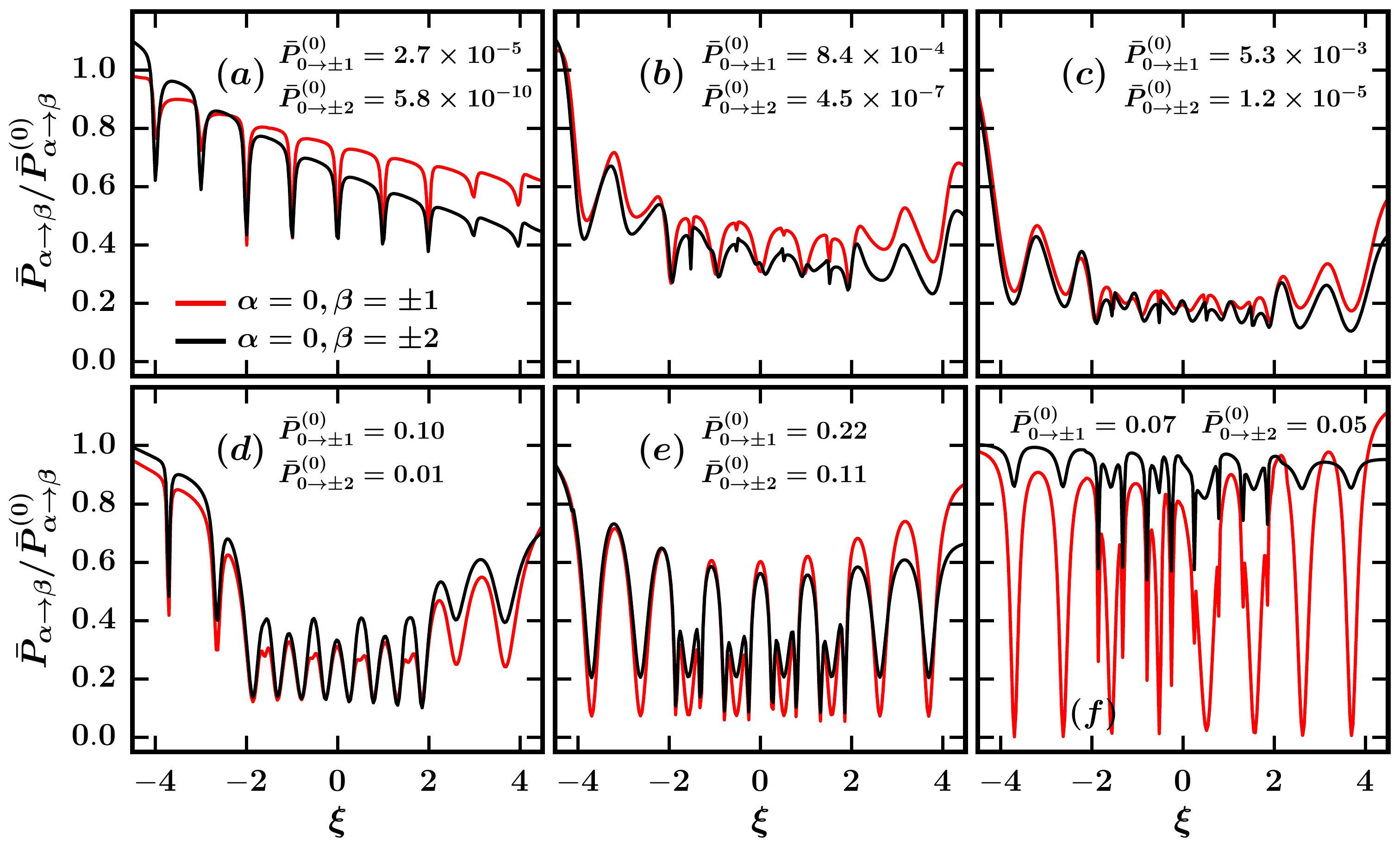}
\caption{\label{fig:brfcomb} Transition probabilities $\bar{P}_{\alpha\rightarrow\beta}$ normalized by the off resonant transition probability $\bar{P}_{\alpha\rightarrow\beta}^{(0)}$ for different rf-field strengths :  
(a) $B_{0}^x$=10 mG, (b) $B_{0}^x$=50 mG, (c) $B_{0}^x$=0.1 G, (d) $B_{0}^x$=0.5 G, (e) $B_{0}^x$=1 G, (f) $B_{0}^x$=5 G. Here, $B_0^y=B_0^z=0$, $\omega_0=2\pi\times\ 2$ MHz, $\omega_r=2\pi\times\ 100$ kHz, $B_q=100\ G\ cm^{-1}$, $N_o=5$.}
\end{figure*}

The mode separation frequency $\omega_r$ in this multi-mode rf-field has been varied and the effect on the eigen-energies has been studied. It has been found that the variation in the $\omega_r$ results in a variation in the spatial profile of the eigen-energies. In Fig. \ref{fig:wr} (a), the variation in the position of the third and forth eigen-energy minima (\textit{i.e.} $\xi_m^{-3}$ and $\xi_m^{-4}$) has been plotted as a function of $\omega_r$. It can be noted that (Fig. \ref{fig:wr} (a)), the position of any energy minimum shows a ``saw-tooth'' type oscillatory behaviour accompanied by sudden jumps at particular values of $\omega_r$ (denoted as $\omega_r^t$). As evident from this figure, the position of an eigen-energy minimum first increases with $\omega_r$, then jumps back to the initial lower value at a particular value of $\omega_r=\omega_r^t$, while the separation between the position of the consecutive minima (\textit{e.g.} $\xi_m^{-3}$ and $\xi_m^{-4}$) remains equal. This oscillatory feature in the position of eigen-energy minimum can be explained as follows. For a $\omega_r$ lying in between two adjacent $\omega_r^t$ values, the energy of the system should increase with an increase in the $\omega_r$. To keep the energy-level separation unchanged with this increase in $\omega_r$, the position of all the eigen-energy minima $\xi_m^k$ gets shifted. At a particular value of $\omega_r=\omega_r^t$, this adiabatic shifting reaches to an extreme when the resonance condition $\omega_L=\omega_0\pm\omega_r$ is achieved in the system, where the $\xi_m^k$ jumps back to the initial value. It is also noted that, the separation between the adjacent $\omega_r^t$ (\textit{i.e.} $\Delta\omega_r^t$) increases linearly which is shown in Fig. \ref{fig:wr} (b). As the resonance condition $\omega_L=\omega_0\pm\omega_r$ is position dependent, the separation $\Delta\omega_r^t$ is expected to vary as a function of $\omega_r$.

One of the key parameters that defines the interaction between the atom and the rf-field is the field strength $B_{0}^x$, as it governs the coupling strength and consequently the transition probability between different states. The dependence of the transition probability $\bar{P}_{\alpha\rightarrow\beta}$ has been studied by varying the rf-field strength $B_{0}^x$ in a wide but experimentally feasible regime, and the results are shown in Fig. \ref{fig:brfcomb}. Apart from the expected sharp variations present near the resonance points, the far resonant saturated values ($\bar{P}_{\alpha\rightarrow\beta}^{(0)}$) also change as a function of the rf-field strength $B_{0}^x$. In Fig. \ref{fig:brfcomb} (a), with very low field strengths ($B_{0}^x$ = 10 mG), not only the single and two photon transition probabilities are vanishingly small, but also the two photon transition probability is 5 orders of magnitude smaller than the single photon transition probability. As the rf-field strength increases, both the transition probabilities increase, but with different rates (shown in Fig. \ref{fig:brfcomb} (b)-(d)). At the field strength 1 G, in Fig. \ref{fig:brfcomb} (e), both the probabilities become very large and the two photon transition probability ($0.11$) is only half of the single photon transition probability (0.22). Finally, more high field strength ($B_{0}^x$ = 5 G in Fig. \ref{fig:brfcomb} (f)) introduces power broadening and destroys the symmetry of the line-shape of the transition probabilities.

This extra ordinary increase in the two photon transition probability, as compared to the single photon transition probability, is a characteristic of interaction of strong polychromatic rf-field with atoms. Due to polychromaticity of the rf-field, avoided crossings are generated either across the states which belong to the same $m$ value (called `primary avoided crossings') or across the states belonging to different $m$ values (called `secondary avoided crossings'). In the primary avoided crossings, the energy levels can not remain arbitrary close to each other due to the conservation of the energy, whereas in case of secondary avoided crossings, no such conservation is required. Hence the energy levels in a secondary avoided crossing can be arbitrary close to each other, which leads to an increase in the transition probabilities due to the resonance enhancements.  

\section{Conclusion}\label{sec:concln}
We have applied a many mode Floquet theory formalism to obtain eigen-energy values of an atomic system interacting with a polychromatic rf-field in presence of a static magnetic field. It is established that this formalism is more promising than the earlier approaches used in this research area. The obtained eigen-energy values show a lattice like periodic spatial variation which is controllable by varying the rf-field parameters such as the separation between the frequencies in the rf-field spectrum. High two photon transition probabilities are also predicted between the atomic states. The results of this work pave the way to realize lattice like atom trapping potentials using polychromatic rf-field and a static magnetic field.

\section*{Acknowledgements}
We acknowledge Charu Mishra for a careful reading of the manuscript. A. Chakraborty acknowledges the financial support by Raja Ramanna Centre for Advanced Technology, Indore under the HBNI programme, Mumbai.

%\bibliographystyle{apsrev}
%\bibliography{Reference}

\end{document}